\documentclass[journal=jacsat,manuscript=article]{achemso}

\usepackage[version=3]{mhchem} % Formula subscripts using \ce{}

\author{Shaojie Hu}
\email{hu.shaojie@phys.kyushu-u.ac.jp} 
\affiliation{Department of Physics, Kyushu University, 744 Motooka, Fukuoka, 819-0395, Japan}

\author{Xiaomin Cui}
\affiliation{Department of Physics, Kyushu University, 744 Motooka, Fukuoka, 819-0395, Japan}

\author{Zengji Yue}
\affiliation{Institute of Photonic Chips, University of Shanghai for Science and Technology, Shanghai, 200093 China.}
\author{Pangpang Wang}
\affiliation{Institute of Systems, Information Technologies and Nanotechnologies (ISIT), Fukuoka, 819-0388, Japan}

\author{Kohei Ohnishi}
\affiliation{Department of Electrical, Electronic and Communication Engineering, Kindai University, 3-4-1 Kowakae, Higashi-Osaka, Osaka 577-8502, Japan}

\author{Shu-Qi Wu}
\affiliation{Institute for Materials Chemistry and Engineering and IRCCS, Kyushu University, 744 Motooka, Nishi-ku, Fukuoka, Japan 819-0395}

\author{Sheng-Qun Su}
\affiliation{Institute for Materials Chemistry and Engineering and IRCCS, Kyushu University, 744 Motooka, Nishi-ku, Fukuoka, Japan 819-0395}

\author{Osamu Sato}
\affiliation{Institute for Materials Chemistry and Engineering and IRCCS, Kyushu University, 744 Motooka, Nishi-ku, Fukuoka, Japan 819-0395}

\author{Sunao Yamada}
\affiliation{Institute of Systems, Information Technologies and Nanotechnologies (ISIT), Fukuoka, 819-0388, Japan}

\author{Takashi Kimura}
\email{t-kimu@phys.kyushu-u.ac.jp}
\affiliation{Department of Physics, Kyushu University, 744 Motooka, Fukuoka, 819-0395, Japan}

\title[An \textsf{achemso} demo]
{Exchange bias induced by spin-glass-like state in Te-rich FeGeTe van der Waals ferromagnet}
\abbreviations{vdW, EB, AHE, FM, AFM,}

\begin{document}

%%%%%%%%%%%%%%%%%%%%%%%%%%%%%%%%%%%%%%%%%%%%%%%%%%%%%%%%%%%%%%%%%%%%%
%% The abstract environment will automatically gobble the contents
%% if an abstract is not used by the target journal.
%%%%%%%%%%%%%%%%%%%%%%%%%%%%%%%%%%%%%%%%%%%%%%%%%%%%%%%%%%%%%%%%%%%%%
\begin{abstract}
  We have experimentally investigated the mechanism of the exchange bias in the 2D van der Waals (vdW) ferromagnets by means of the anomalous Hall effect (AHE) together with the dynamical magnetization property. 
The temperature dependence of the AC susceptibility with its frequency response indicates a glassy transition of the magnetic property for the Te-rich FeGeTe vdW ferromagnet. We also found that the irreversible temperature dependence in the anomalous Hall voltage follows the Almeida-Thouless line.  Moreover, the freezing temperature of the spin-glass-like phase is found to correlate with the disappearance temperature of the exchange bias. These important signatures suggest that the emergence of magnetic exchange bias in the 2D van der Waals ferromagnets is induced by the presence of the spin-glass-like state in FeGeTe. The unprecedented insights gained from these findings shed light on the underlying principles governing exchange bias in vdW ferromagnets, contributing to the advancement of our understanding in this field.
\end{abstract}

%%%%%%%%%%%%%%%%%%%%%%%%%%%%%%%%%%%%%%%%%%%%%%%%%%%%%%%%%%%%%%%%%%%%%
%% Start the main part of the manuscript here.
%%%%%%%%%%%%%%%%%%%%%%%%%%%%%%%%%%%%%%%%%%%%%%%%%%%%%%%%%%%%%%%%%%%%%
\section{Introduction}
In recent years, 2D vdW materials have garnered significant interest for their unique properties, which are not readily attainable in their bulk crystal forms. Notably, a recent and remarkable development in this domain is the discovery of 2D vdW ferromagnets. These materials include intrinsic ferromagnetic (FM) and anti-ferromagnetic (AFM) orders\cite{gong2017discovery, huang2017layer, deng2018gate, fei2018two-dimensional}, expanding the scope of research and applications in 2D materials. The weak interlayer coupling in 2D vdW magnets facilitates their easy exfoliation and stacking. This structural characteristic makes these 2D vdW magnetic materials an ideal platform for the in-depth study of spin coupling phenomena. 
Among the spin coupling effects, the exchange bias (EB) effect, or exchange anisotropy, is particularly notable.\cite{Meiklejohn1956,nogues2005exchange,hu2012ferromagnetic,chang2023field}
This interfacial magnetic coupling effect plays an important role 
in designing a range of sophisticated spintronic devices. Though it has been extensively studied in the heterojunctions with 2D van der Waals ferromagnets, such as $\rm CrI_3/Fe_3GeTe_2$, $\rm Fe_3GeTe_2/FePS_3$, $\rm Fe_3GeTe_2/MnPX$, $\rm Fe_3GeTe_2/IrMn$, $\rm Fe_3GeTe_2/CrOCl$, $\rm Fe_3GeTe_2/CoPc$. \cite{Zhu2020,Hu2020,Zhang2020Proximity,Zhang2021Exchange,Dai2021Enhancement,zhang2022tuning,jo2022exchange}. 
Interestingly, the exchange bias effect has been observed in standalone structures, such as the van der Waals ferromagnet $\rm Fe_{3-x}GeTe_2$ (FGT) nano-flake, without the need for hybrid junctions.\cite{Zheng2020gate,Gweon2021,hu2022positive,liu2022emergent} 
Currently, a consistent explanation for the origin of the exchange bias in this vdW ferromagnet remains elusive. 
The exchange bias effects in vdW FGT nanoflakes can be modulated through gate-induced proton intercalation.\cite{Zheng2020gate,wang2023sign}
The induction of the EB effect in vdW FGT can also be attributed to the oxidation of its surface, which leads to the formation of AFM states in the oxidized layer\cite{Gweon2021}. Additionally, positive exchange-biased behavior has been observed in vdW FGT nanoflakes\cite{hu2022positive}.
Correspondingly, the origin of this phenomenon has become a subject of considerable debate, particularly the unclear role of some disordered spin and frustration either at the interface or throughout the entire system. 

Several reports claimed the coexistence of the AFM and FM in FGT nanoflakes.\cite{Yi2017, Lachman2020,hu2022positive} 
Experimental findings have shown that competing and coexisting FM and AFM states are present in the temperature range of 152-214 K \cite{Yi2017}. Moreover, in our recent study, the observed frustrated positive exchange bias has been explained by the coexistence of local anti-ferromagnetic (AFM) and ferromagnetic (FM) states \cite{hu2022positive}.  
The coexistence of local antiferromagnetic spin fluctuation and itinerant ferromagnetic spin has also been confirmed  through neutron scattering and thermodynamic measurements in 
$\rm Fe_{3-x}GeTe_2$ with less rich Fe.\cite{bai2022antiferromagnetic}
In addition, local antiferromagnetic coupling results from Te substitution at Ge sites and Fe-intercalation, as identified by atomic-resolution EDS mapping and DFT calculation.\cite{wu2023fe} These findings suggest that the atomic composition plays a crucial role in forming local AFM states in such vdW ferromagnets.
In view of the coexistence of the AFM and FM phase in the FGT,  it may be a chance to form the spin-glass-like phase in this vdW materials because of a largely random-looking mixture of ferromagnet and antiferromagnet.\cite{weissman1993} 
The spin-glass could be characterized to demonstrate some low-temperature spin freezing and glassy dynamics\cite{fischer1993spin, goremychkin2008,mydosh2015spin}. 
The exchange bias effect, induced by spin-glass, has been observed in numerous hybrid systems.\cite{ali2007exchange,usadel2009exchange,maniv2021exchange}. 
However, this effect has yet to be reported in vdW ferromagnets, indicating a potential area for exploration in these unique materials.
In this paper, a systemic study of the cooling field and temperature dependence of anomalous Hall effect reveals the existence of the spin-glass-like properties and their relation to the exchange bias effect in rich-Te vdW FGT.
\section{Results and discussion}

  \begin{figure*}[ht]
	\centering
	\includegraphics[width=6.3in]{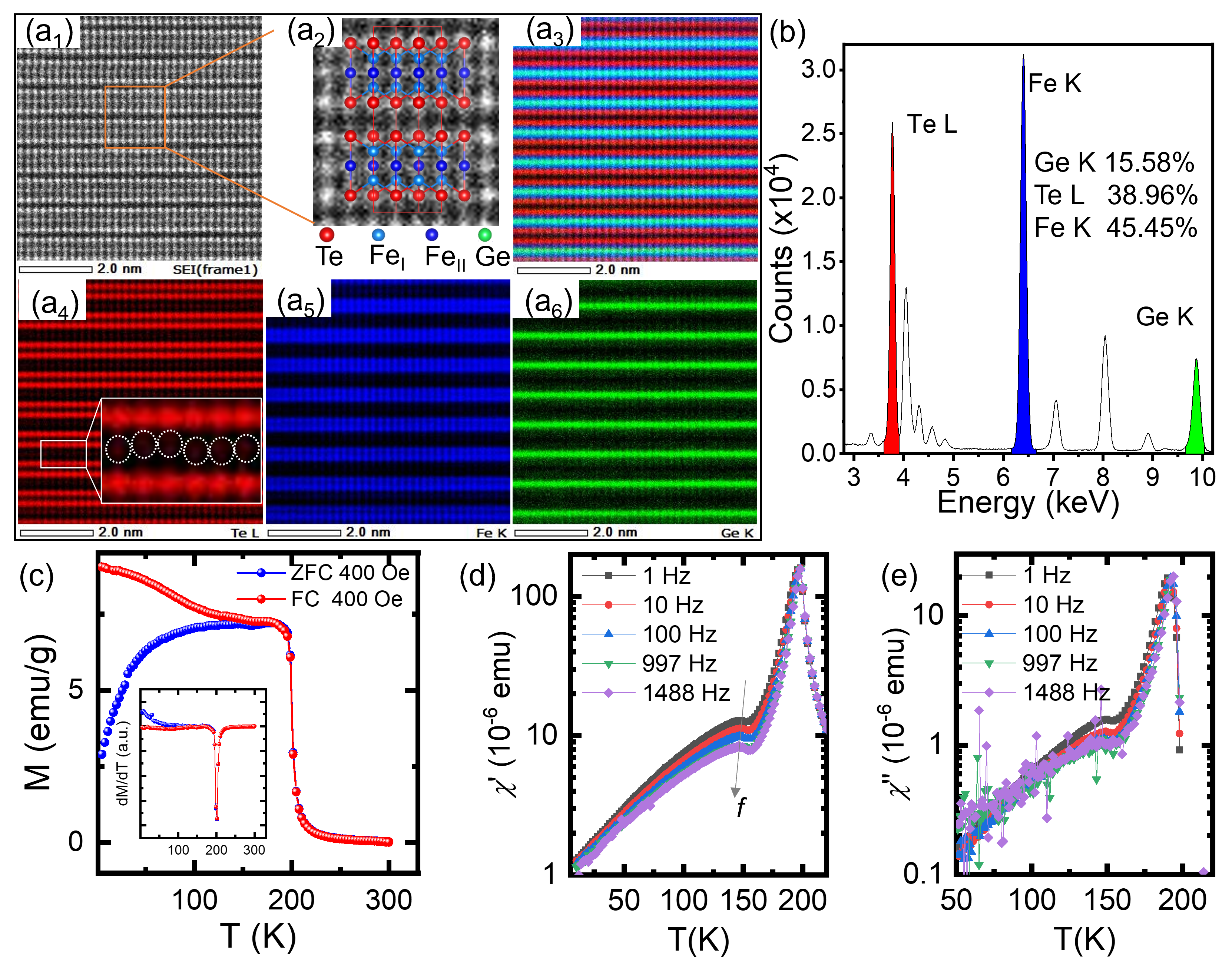}	
	\caption{
		($\rm a_1$) - ($\rm a_6$) The HAADF-STEM images showcase the (100) plane of FGT, accompanied by corresponding EDS mapping images highlighting the distribution of Te, Fe, and Ge atoms. The higher magnification of the STEM image ($\rm a_2$) shows an almost identical atomic structure. The red, light-blue/blue, and green balls represent Te, Fe and Ge atoms, respectively.  2$\rm Fe^I$ and $\rm Fe^{II}$, which are sandwiched by Te atoms, indicate the two distinct Fe sites with +3 and +2 valence states, respectively.  Inset of ($\rm a_4$) illustrates some Te atoms (marked by white dot circles ) occupying sites typically held by Fe atoms.
        (b) The EDS analysis verifies the atomic ratios of Fe, Ge, and Te are 45.45 \%, 15.58 \% and 38.96 \%, respectively. The stoichiometric composition of FGT is Fe : Ge : Te =  2.73: 0.93 : 2.34 by considering the 6 atoms in one unit cell.
	(c) The field-cooled (FC) and zero field-cooled (ZFC) thermal magnetization curves with a 400 Oe magnetic field applied along the c axis of the bulk crystal. The Curie temperature is about 200K ($\rm T _C \approx 200K$). Insert: the derivative magnetization $dM/dT$ vs $T$.
    The real (d) and imaginary (e) components of temperature-dependent AC susceptibility were measured under zero DC external field and an oscillating field of 3 Oe for various frequencies. 
}
\end{figure*}

\subsection{Characterization of FGT single crystal }
  
  Fig. 1 shows the high-resolution high-angle annular dark-field scanning transmission electron microscopy (HAADF-STEM) images of FGT in (100) plane and energy-dispersive X-ray spectroscopy (EDS) mapping images highlighting the distribution of Te, Fe, and Ge atoms. The higher magnification of the STEM image ($\rm a_2$) shows an almost identical atomic structure. The red, light-blue/blue, and green balls represent Fe, Ge, and Te atoms, respectively.  2$\rm Fe^I$ and $\rm Fe^{II}$, which are sandwiched by Te atoms, indicate the two distinct Fe sites with +3 and +2 valence states, respectively.  The inset of ($\rm a_4$) illustrates some Te atoms (marked by white dot circles ) occupying sites typically held by top $\rm Fe^{I}$ and $\rm Fe^{II}$/Ge atoms. This finding indicates that the occupying sites are not all uniformly distributed in the systems. To confirm the ratio of the total occupying site, we analyse the EDS spectrum as shown in Fig. 1(b) and confirm the stoichiometry of $\rm Fe_{2.79}Ge_{0.98}Te_{2.23}$.  
  Such results indicate that Te substitutes about the 7 \% of Fe and 2 \% of Ge sites.  The substitution of $\rm Fe^{I}$ or $\rm Fe^{II}$ are not uniformly distributed in FGT shown in Fig.1  ($\rm a_4$). So, the Te substitution could break the spacial inversion symmetry in some local regions of this system. The conducted electrons could be strongly scattered by local spin momentum due to the Kondo effect\cite{bao2022neutron,zhao2021kondo}.
  This alteration could affect the Ruderman–Kittel–Kasuya–Yosida (RKKY) interaction, potentially inducing local AFM coupling in this itinerant ferromagnet.
 
 The field-cooled (FC) and zero-field-cooled (ZFC) thermal magnetization curves are shown in Fig.1(c), with a 400 Oe magnetic field applied along the c axis of the bulk crystal. The Curie temperature (T$ \rm _C $) is about 200K, where a sharp decrease appears in the first derivative magnetization curve shown inset.  The T$ \rm _C $ of our FGT sample is slightly lower than the general Curie temperature 220 K, because of the less rich amount of Fe.\cite{doi:10.1002/ejic.200501020,may2016magnetic,Liu2017Wafer,liu2017critical,wu2023fe}  
  Fig.1(c) shows a consistent increase in the magnetic moment of FGT as the temperature decreases below 150 K in the field-cooled (FC) curve. Although many published studies report similar findings, they often lack sufficient explanation.\cite{Yi2017,wang2017anisotropic,drachuck2018effect,Tian2019} 
  To further examine the abnormal magnetic behavior, the temperature-dependent AC susceptibility of the bulk crystal was measured under zero DC external field with various frequencies, as shown in Fig.1(d) and (e). During this measurement, the oscillated AC field of 3 Oe is applied in the out-of-plane direction.
  There are two distinct stable magnetic transitions around 200 K and 150 K. The 200 K is the Curie temperature, which is related to the magnetic phase to the paramagnetic phase.  However, the lower transition temperature indicated there may be a new magnetic phase below 150 K.  Compared to the paramagnetic phase transition peak, the broadening of the peak around 150 K denotes a considerable range in the distribution of relaxation times. A small frequency dependence of the transition peak around 150 K indicates a glassy transition. 
  The Mydosh parameter\cite{mulder1981susceptibility,binder1986spin,mydosh2015spin} is about $\Delta T/(T_{sg} log_{10}(\Delta f)\ \approx$ 0.0063, which is used to qualitatively identify the type of glassy dynamics associated with a transition within the range expected of a spin-glass transition.
  Typically, a new magnetic phase transition impacts spin-related transport properties. 
  %Since FGT is an efficient ferromagnetic conductor, its electrical transport properties serve as a vital means to analyze and comprehend the magnetic phase transitions in nano-flakes. Notably, the Anomalous Hall Effect (AHE) is an effective method for acquiring information about the ferromagnetic characteristics of vdW ferromagnets.
  
 \begin{figure}[!ht]
 	\centering
 	\includegraphics[width=5in]{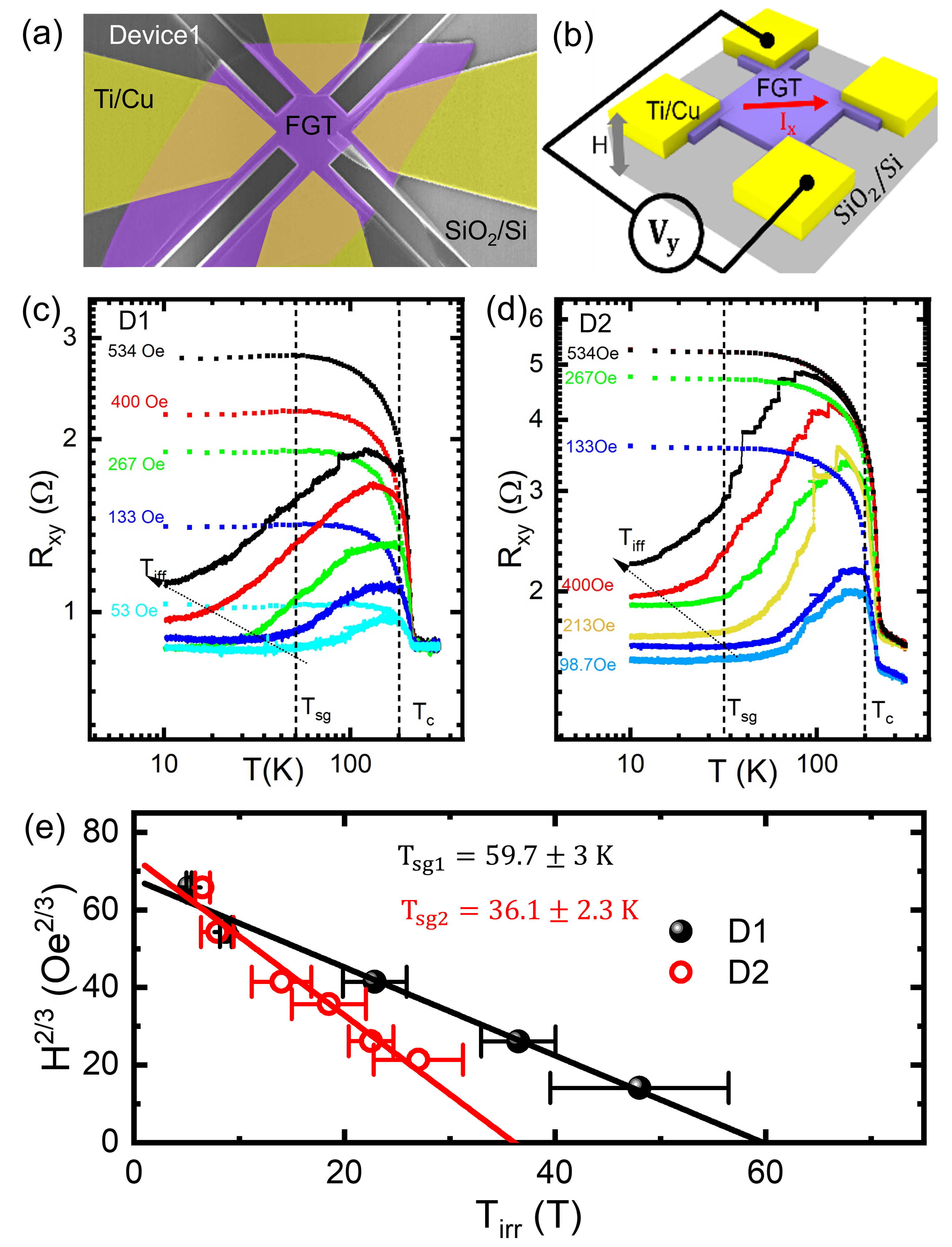}	
 	\caption{(a) Scanning electron microscopy (SEM) image of one fabricated with a cloverleaf structure.
  (b) The schematic diagram illustrates the design of the cloverleaf-structured device, along with the configuration for the Anomalous Hall Effect measurement. The width of the square is about $\rm 2\mu m$. 
  (c,d) The temperature dependence of the  $\rm R_{xy}$ is obtained by increasing the temperature from zero-field-cooled (ZFC) and field-cooled (FC) for the various steady fields for D1 and D2. The dot (solid) lines are representing the FC (ZFC) results. The irreversibility temperature $\rm T_{irr}$ is the starting point for increasing $\rm R_{xy}$ in the ZFC curves. 
  (e) The irreversibility temperature dependence of 2/3 power of the magnetic field power of the magnetic fields for D1 (solid black) and D2 (red hollow dot). The solid lines represent the linear fitting of AT  function.}
 \end{figure}

 \subsection{Spin-glass-like behavior of FGT}
The scanning electron microscopy (SEM) image of device 1 (D1) with a cloverleaf structure is shown in Fig.2(a). 
 Fig.2 (b) shows the schematic measurement configuration of the AHE effect. 
First, the temperature dependence of the  $R_{\rm xy}=V_{y}/I_x$ is recorded by increasing the temperature for zero-field-cooled (ZFC) and field-cooled (FC) with the various out-of-plane steady fields for D1 and D2, as shown in Fig.2 (c) and (d). All $R_{\rm xy}$ significantly reduce when the temperature is over Curie temperature.  
It should be noted that the $R_{\rm xy}$ is slightly declining with a reduction of temperature around 60 K for the higher  FC curves of D1. Such a reduction of $R_{\rm xy}$ should relate to some new phase transition, which should share the same mechanism with the second transition temperature (around 150 K) in bulk.
To better understand the behavior of these properties, we can also confirm that the $R_{\rm xy}$ is almost flat below the specific temperature at the low magnetic field for the ZFC curves. It means the spin should be frozen below this temperature. A pronounced peak can also be seen in the ZFC curves of $R_{\rm xy}$.  All such features look like random-field-induced spin-glass-like behavior. From the mean-field theory, there are also critical lines, which are the variables of temperature and magnetic fields in the spin-glass systems. 

In spin systems by mean-field theory, the de Almeida and Thouless predicted the so-called AT line is usually defined as \cite{binder1986spin,Gruyters2005} 
\begin{equation}\label{ATline}
H_{\rm AT}/\Delta J \propto (1-T_{\rm irr}/T_{\rm sg})^{3/2}
\end{equation}
where the irreversible temperature $T_{\rm irr}$ is the starting point for increasing $R_{\rm xy}$ in ZFC curves. $T_{\rm sg}$ is the zero-field spin-glass freezing temperature. $\Delta J$ is the width of the distribution of exchange interactions. 

The irreversible temperature dependence of 2/3 power of the magnetic field shows the strongly linear relation in Fig.2(e). By fitting the Eq.(1),  the zero-field spin-glass freezing temperature $T_{\rm sg}$ of D1 and D2 are estimated as $\rm 59.7\pm 3$ K and $\rm 36.1\pm 2.3$ K, respectively. 
The variation in freezing temperature could be attributed to differences in the strength of anisotropy energy, which can be influenced by the surface oxidation.\cite{montenegro1991random,nowak1991diluted,han1992relaxation} 
When the temperature is below $T_{\rm sg}$, the localized spins are locked into the random distribution.% because of the direct interaction between the Fe intercalation/defect or hole doping. 
If the temperature rises above $T_{\rm sg}$, the thermal energy of the localized spins is sufficient to overcome the RKKY or Kondo interactions induced ordering influences. It becomes the frustrating ferromagnetic state at this temperature range  $T_{\rm sg} < T<T_{\rm C}$, where we can also see some significant jumping behavior for the ZFC curves. %This behavior may be induced by the pinned local ferromagnetic domains.
When the temperature exceeds $T_{\rm C}$ = 200 K,  $R_{\rm xy}$ becomes zero because of the paramagnetic state. 

\begin{figure}[ht]
	\centering
	\includegraphics[width=6.3in]{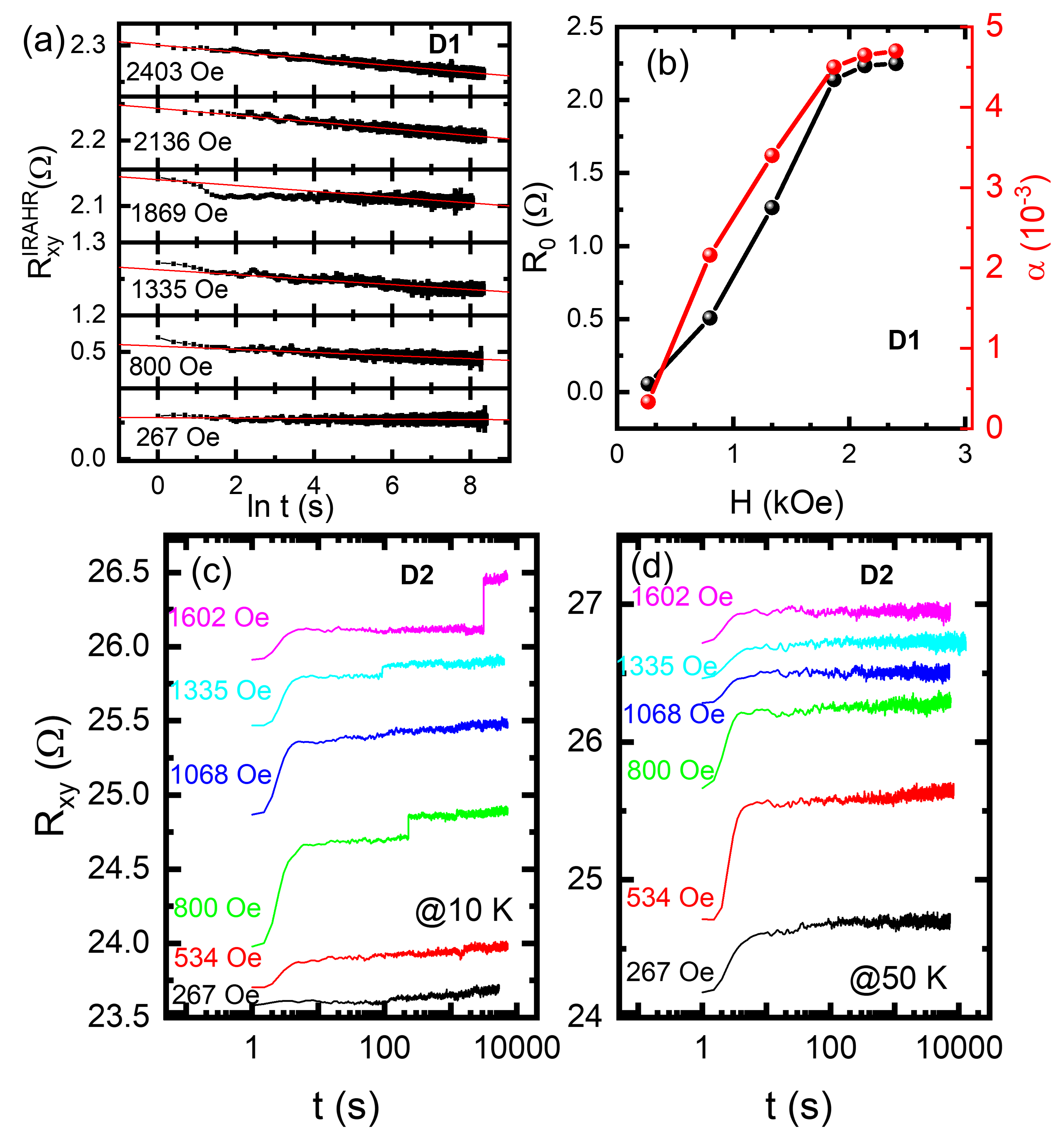}	
	\caption{
	 (a)  Isothermal remanent anomalous Hall resistance ($R_{\rm xy}^{\rm IRAHR}$) of device 1 is plotted as a function of time t for various external field at 10 K . The solid lines are fits to the experimental data according to Eq.2.  
    (b) The external field dependence of fitting parameters $R_{\rm xy}^{\rm IRAHR}(t=0)$ and $\alpha$.
    The time-dependent $R_{\rm xy}$ under a constant magnetic field, both below (c) and above (d) the spin glass transition temperature ($T_{\rm sg}$) of device 2,
	}
\end{figure}

In addition, the spin-glass-like system could be studied by evaluating the time dependence of isothermal remanent magnetization (IRM). The measured anomalous Hall resistance $R_{\rm xy}$ is proportional to the magnetization of the materials. Therefore, we measured the isothermal remanent anomalous Hall resistance (IRAHR) $R_{\rm xy}^{\rm IRAHR}$ to understand the IRM properties indirectly. %Fig.3  presents the time dependence of $R_{\rm xy}^{\rm IRAHR}$.% under various external magnetic fields. 
Isothermal remanent anomalous Hall resistance was measured under the ZFC process from 300 K to 10 K and acted on an external magnetic field for 600 s. $R_{\rm xy}^{\rm IRAHR}$ was recorded after switching off the magnetic field. The gradual decay of $R_{\rm xy}^{\rm IRAHR}$ aligns with the spin-glass-like behavior observed in the diluted Ising antiferromagnetic system.\cite{nowak1990slow,montenegro1991random,nowak1991diluted,han1992relaxation,mydosh2015spin} 
Fig.3 (a) shows the experimental data $R_{\rm xy}^{\rm IRAHR}$ fitted according to the formula \cite{binder1986spin} 
\begin{equation}\label{IRAHR}
R_{\rm xy}^{\rm IRAHR}(t)=R_{0}-\alpha ln(t)
\end{equation}
where $\rm \alpha \ and \ R_{0}$ are the fitting parameters.  $\rm \alpha$ is related to the magnetic viscosity.\cite{guy1978spin} Clearly, Figure 3(b) demonstrates that both fitting parameters escalate before plateauing with an external magnetic field up to 2.4 kOe, a characteristic typically observed in the spin-glass-like system. 
To gain a more comprehensive understanding of spin-glass-like behavior, we also conducted measurements of time-dependent $R_{\rm xy}$ with a continuously applied magnetic field, both below and above the spin-glass transition temperature ($T_{\rm sg}$) of D2, as illustrated in Fig.3(c,d). Prior to these measurements, the device was subjected to a cooling process from 300 K to either 10 K or 50 K without a magnetic field.
Initially, a magnetic field of 267 Oe was implemented, with $R_{\rm xy}$ recorded for an approximate duration of 6000 s. Subsequently, we augmented the applied field to other magnitudes and resumed the recording of $R_{\rm xy}$. Notably, a significant surge in $R_{\rm xy}$ was observed at distinct magnetic fields for 10 K. However, this sharp rise in $R_{\rm xy}$ behavior was absent at 50 K, implying the existence of a metastable state at 10 K. Following a specified waiting period, the metastable state transitioned into an equilibrium state, facilitated by the relaxation of the spin-glass-like. Consequently, no metastable state was observed above the spin-glass temperature ($T_{\rm sg}$ $\approx$ 36.1 K). 

\begin{figure}[ht]
	\centering
	\includegraphics[width=6in]{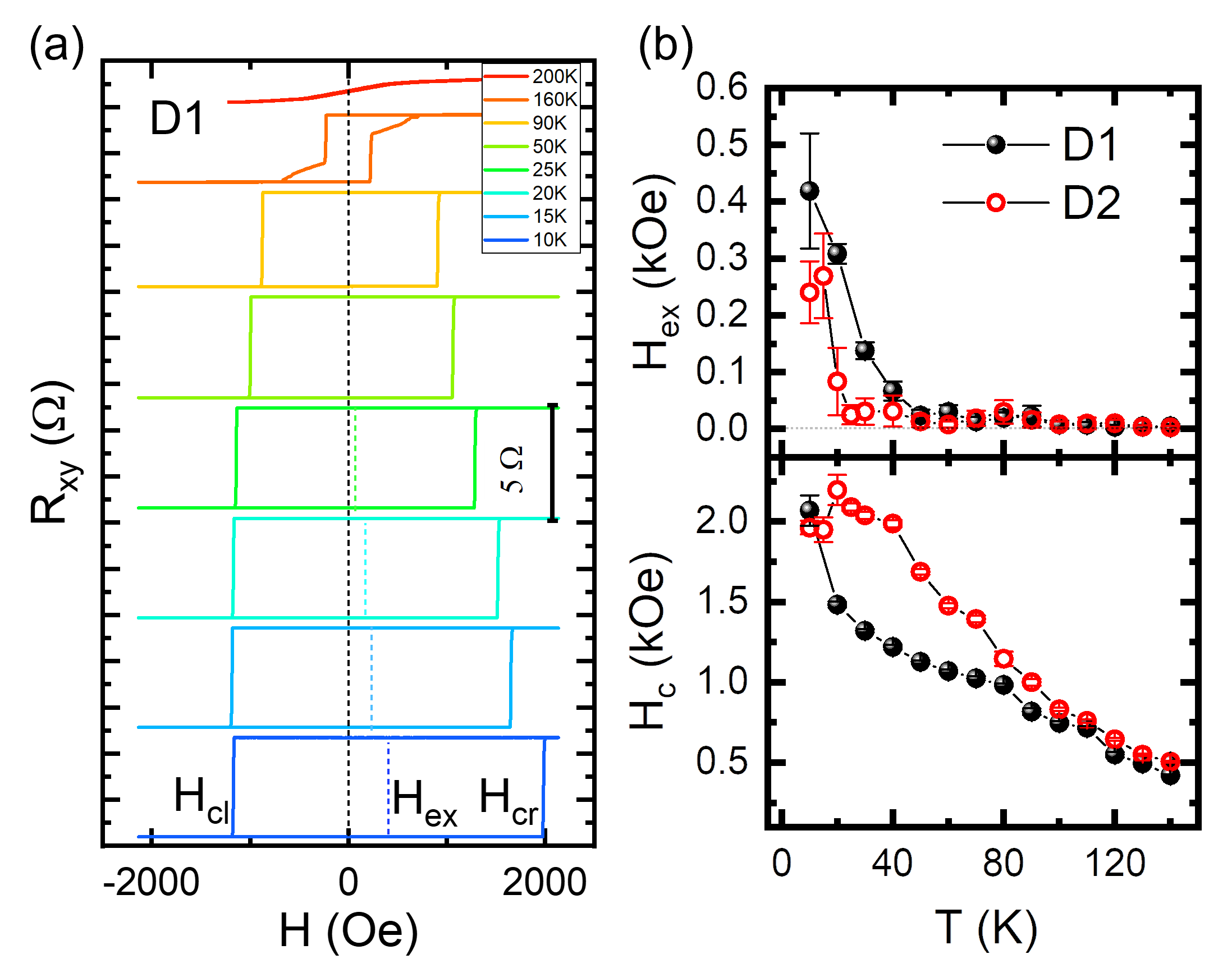}	
	\caption{ 
 (a) AHE signals are obtained by measuring $R_{\rm xy}$ with sweeping out-off plane magnetic field from -3000 Oe to +3000 Oe and sweeping back to -3000 Oe. The $H_{\rm cl}$ and $H_{\rm cr}$ are the negative and positive switching field, respectively.
The exchange biased field is defined as $H_{\rm ex}=(H_{\rm cl}+H_{\rm cr})/2$. The coercivity field is  $H_{\rm c}=(H_{\rm cr}-H_{\rm cl})/2$.
(b) The temperature dependence of the $H_{\rm ex}$ and $H_{\rm c}$ . The devices were initially cooled to 2.4 K with a magnetic field of 2400 Oe. The black solid dots and red hollow dots represent the results for D1 and D2, respectively.}
\label{Fig4}
\end{figure}

\subsection{Exchange bias effect of FGT}
Then, the temperature dependence of the anomalous Hall effect was investigated. This was accomplished by sweeping the magnetic field from -3000 Oe to +3000 Oe and then reversing the sweep back to -3000 Oe. Data acquisition was conducted subsequent to initial field cooling at 2400 Oe down to 2.4 K. Following this, anomalous Hall effect (AHE) curves were meticulously recorded over 30 cycles for each temperature setting, subsequent to a steady increase to the predetermined temperature values.
Figure 4 (a) shows the anomalous Hall signals from the temperature range of 10 K to 200 K for D1.
Here, we defined the negative and positive switching fields as $H_{\rm cl}$ and $H_{\rm cr}$, respectively. The coercivity field is $H_{\rm c}=(H_{\rm cr}-H_{\rm cl})/2$.
The exchange biased field is defined as $H_{\rm ex}=(H_{\rm cl}+H_{\rm cr})/2$, which is the absolute offset of loops along the field axis. 
Fig.4 (a) shows the clearly positive exchange bias effect at lower temperatures, which quickly disappears with increasing temperature.

To better understand this behavior, we also plot the temperature dependence of the exchange bias field for two different devices in Fig.4(b). Both the averaged amplitudes of $H_{\rm ex}$ and $H_{\rm c}$ are obtained from 30 $R_{\rm xy}$-H loops. The $H_{\rm ex}$ of D1 becomes to be zero at about 60 K, which is defined as the disappearance temperature of $H_{\rm ex}$.
For the D2, the disappearance temperature of $H_{\rm ex}$ is around 30 K. Here, we observe that all the disappearance temperatures are associated with spin glass transition temperatures.
In this system, Te substitution-induced local AFM coupling is dispersed randomly, thereby constituting a spin-glass-like system. Consequently, an analysis of the exchange bias effect in the entire system cannot be accurately made by focusing solely on one local antiferromagnetic property. 
The total surface antiferromagnetic/ferromagnetic coupling energy density in the system can thus be inferred to be linked to the spin-glass temperature. Within the scope of the entire system, the exchange bias field bears a strong relationship with the total surface antiferromagnetic/ferromagnetic coupling energy density and the anisotropy energy density of the ferromagnets, as proposed in the Meiklejohn-Bean model \cite{meiklejohn1957new}. It is understood that the magnitude of $H_{\rm c}$ is mainly derived from the magnetic anisotropy energy density.
Figure 4 (b) presents a typical thermal activation behavior: a decreasing trend in $H_{\rm c}$  with an increase in temperature. However, the decreasing tendency is a bit different for the two devices. Specifically, D1 exhibits a significantly lower value of $H_{\rm c}$  below 80 K compared to D2, indicating a greater anisotropy energy density for D2 at lower temperatures. But the $H_{\rm c}$ is a bit lower. 
As a consequence, this could lead to a reduction in the spin-glass transition temperature. The increase in the anisotropy energy density, coupled with a decrease in the total surface antiferromagnetic/ferromagnetic coupling energy density, could potentially yield a lower exchange bias field and spin-glass-like transition temperature in D2.
%For D2, the total surface antiferromagnetic/ferromagnetic coupling energy density might be diminished quickly with increasing temperature due to substrate surface scattering. 

\section{Conclusion}
In summary, the temperature-dependent AC susceptibility results of the FGT bulk crystal indicate there are two distinct stable magnetic transitions around 200 K and 150 K. A small frequency dependence of the lower transition temperature peak indicates a glassy transition.
The study systematically investigated the temperature dependence of the anomalous Hall effect in Te-rich FGT flakes, revealing clear spin-glass-like behavior. Notably, the disappearance temperature of the exchange bias aligns with the freezing temperature of the spin-glass-like phase, indicating that the spin-glass-like state is likely responsible for the emergence of magnetic exchange bias in FGT flakes. These findings provide unprecedented insights into the underlying mechanisms governing exchange bias in vdW ferromagnets, enriching our understanding of this field.

\section{Experimental method}

\subsection*{Materials and characterization}
The single crystals of FGT were produced by utilizing the chemical vapor transport (CVT) method with iodine ($\rm I_2$) as a transport agent. As the starting ingredients, high-purity stoichiometric quantity powders (1 g) of Fe, Ge, and Te, along with 10 mg/ml iodine, are sealed in a quartz tube. The bulk crystal was grown in a two-zone furnace between 750 $^{\circ}$C (source) and 700 $^{\circ}$C (sink) for one week. 
The TEM samples were prepared using the focused ion beam technique (Triple Beam FIB-SEM-Ar Grand MI4000L from Hitachi High Technologies). 
The double Cs-corrected scanning transmission electron microscopy (STEM, Grand JEM-ARM300F) was applied to analyse the atomic structure. The microscopy was equipped with a cold field-emission gun and operated at the accelerating voltage of 300 kV. The energy-dispersive X-ray spectroscopy mapping was applied for chemical analysis and the element distribution study.
The temperature dependence of magnetization (M-T) curve and AC susceptibility were measured using a superconducting quantum interference device (SQUID).

\subsection*{Device fabrication and evaluation} 
The thin flakes of FeGeTe were mechanically exfoliated from the bulk crystals and transferred onto SiO$_2$/Si substrate by scotch tape in a glove box filled with argon. The photo-resist was then spin-coated onto the substrate to prevent quick oxidation of the FGT flakes. 
The electrodes of Ti/Cu were fabricated by using electron beam lithography (EBL) system (ELIONIX ELS-7800) and a wet lift-off process. The 5 nm Ti was deposited in the ultra-high vacuum electron beam evaporator to obtain good adhesion. Then 100 nm Cu was deposited by using a Joule evaporator in the same chamber without breaking the vacuum. After that, we used argon ion-beam milling to cut away the extra part of the thin FGT flake to make the standard cloverleaf shape. 
The anomalous Hall effect was measured in a cryostat with a temperature-control system.  The anomalous Hall voltage was detected with the lock-in amplifier by applying a low AC current of 173 Hz. The magnetic field $\rm H$ is sweeping out of the FGT surface plane, as shown in Fig.2(b). The magnetic field sweep rate is about 13  Oe/s, and the time constant is 300 ms for reading data. The AHE signals are evaluated in a temperature range of 10 K to 300 K. 
The thicknesses of the FGT flakes in D1 and D2 are approximately 47 nm and 27 nm, respectively. These measurements were obtained using transmission electron microscopy, which was conducted after completing all electrical assessments for each device.

%%%%%%%%%%%%%%%%%%%%%%%%%%%%%%%%%%%%%%%%%%%%%%%%%%%%%%%%%%%%%%%%%%%%%
%% The "Acknowledgement" section can be given in all manuscript
%% classes.  This should be given within the "acknowledgement"
%% environment, which will make the correct section or running title.
%%%%%%%%%%%%%%%%%%%%%%%%%%%%%%%%%%%%%%%%%%%%%%%%%%%%%%%%%%%%%%%%%%%%%
\begin{acknowledgement}
This work is partially supported by National JSPS Program for Grant-in-Aid for Scientific Research (S)(21H05021), and Challenging Exploratory Research (17H06227) and JST CREST (JPMJCR18J1). The authors thank the Analysis Center of Fukuoka Industry-Academia Symphonicity for TEM sample preparation by using focused ion beam (FIB). The high-resolution STEM observation was supported by “Advanced Research Infrastructure for Materials and Nanotechnology in Japan (ARIM)” of the Ministry of Education, Culture, Sports, Science and Technology (MEXT). Grant Number JPMXP1222KU0013.

\end{acknowledgement}

%%%%%%%%%%%%%%%%%%%%%%%%%%%%%%%%%%%%%%%%%%%%%%%%%%%%%%%%%%%%%%%%%%%%%
%% The same is true for Supporting Information, which should use the
%% suppinfo environment.
%%%%%%%%%%%%%%%%%%%%%%%%%%%%%%%%%%%%%%%%%%%%%%%%%%%%%%%%%%%%%%%%%%%%%
%\begin{suppinfo}

%This will usually read something like: ``Experimental procedures and
%characterization data for all new compounds. The class will
%automatically add a sentence pointing to the information on-line:
%\end{suppinfo}

%%%%%%%%%%%%%%%%%%%%%%%%%%%%%%%%%%%%%%%%%%%%%%%%%%%%%%%%%%%%%%%%%%%%%
%% The appropriate \bibliography command should be placed here.
%% Notice that the class file automatically sets \bibliographystyle
%% and also names the section correctly.
%%%%%%%%%%%%%%%%%%%%%%%%%%%%%%%%%%%%%%%%%%%%%%%%%%%%%%%%%%%%%%%%%%%%%
\bibliography{FGT}

\end{document}